\documentclass[11pt,twoside]{article}
\usepackage{asp2014}

\aspSuppressVolSlug
\resetcounters

\begin{document}

\title{White dwarf research with Gaia}
\author{Stefan Jordan}
\affil{Astronomisches Rechen-Institut am Zentrum f\"ur Astronomie der \\
Universit\"at Heidelberg, Germany; \email{jordan@ari.uni-heidelberg.de}}

\paperauthor{Stefan Jordan}{jordan@ari.uni-heidelberg.de}{0000-0001-6316-6831}{Universit\"at Heidelberg}{Astronomisches Rechen-Institut am Zentrum f\"ur Astronomie der Universit\"at Heidelberg}{Heidelberg}{}{D-69120}{Germany}
\begin{abstract}
The results of the Gaia mission will have tremendous influence on many topics in white dwarf research. In this paper the current status of the Gaia mission is described. At the end a short outlook on the release scenario and the expected accuracy of the Gaia data is provided.
\end{abstract}

\section{The Gaia mission}
On December 19, 2013, the ESA astrometry satellite Gaia was launched with a Russian Soyuz-Fregat rocked from Sinnamary, close to Kurou in French Guiana. Since January 14, 2014, Gaia is in an orbit around the Langrage point L2 of the Sun-Earth system. 
After a commissioning phase, which ended in July 2014, Gaia started its routine phase. 

Gaia is equipped with two telescopes with rectangular main mirrors of size $145$\, cm $
\times 50$\,cm aperture and a focal length of 35\,m, which point at two different regions in the sky 106.5 degrees apart \cite[see e.g. ][]{GaiaTechnique}. 

With its two telescopes Gaia shall scan the complete sky several times in different directions
during its nominal mission of five years. All point sources brighter than magnitude 20 are automatically detected and registered by Gaia's camera consisting of 106 CCD detectors. A  small "window" of CCD pixels  around each stellar image is  downlinked  to ground during. The actual astrometric measurements consist of determining the time when the centroid of a stellar image is detected by the CCDs which are  operated in time delay and integration (TDI) mode. This means that the speed of the readout of the CCDs must correspond to the speed of the stellar images moving over the focal plane due to the rotation of the satellite.

Low-resolution spectra ($\Delta \lambda/\lambda<100$) for all $V<20$ stars are provided by a blue (3300-6800\AA) and red (6400-10500\AA)  photometer (BP and RP). These measurements are used for the astrometric chromaticity correction as well as for the classification and characterization of the observed stars. 
Moreover, high-resolution ($\Delta \lambda/\lambda\approx 11000$) near-infrared spectra (8450\, \AA-8720\AA) are taken by the radial velocity spectrograph RVS for the brighter stars ($V<16$) . 

On the average every star is going through one of the two fields of view 75 times during the nominal mission.  For each transit up to 10 astrometric measurements  are taken together with white band photometry between 3300-10500\AA;  additionally one BP, one RP, and three RVS spectra are gathered. 

In total about $10^{12}$ single astrometric measurements are performed during a mission of five years. Assuming one billion stars down to 20$^{th}$ magnitude  5 billion astrometric parameters (for each single star two position coordinates, two values for the proper motion, and one for the parallax) have to be determined plus additional orbital parameters for stars in multiple systems.
This is done with a block-iterative  least-squares solution  \citep[Astrometric Global Iterative Solution, AGIS,][]{AGIS} also taking into account many hundred million unknowns describing the attitude of the satellite as a function of time as well as 
other geometric parameters.

\section{White dwarfs}
The results of the Gaia mission will have great impact on practically all fields of astronomy. This is   in particular true for white dwarf research and a large fraction of all talks during this conference have been mentioning the necessity to use Gaia data. 

It is expected that Gaia will discover of the order of 400,000 new white dwarfs \citep{Torres2005,Jordan2007}. The discovery will be possible by combining the photometric data with the astrometric information. In the ZZ Ceti range around $T_{\rm eff}\approx 12000$\,K, where the Balmer lines are strongest, H$_\alpha$,H$_\beta$, and  H$_\gamma$ can directly be detected as broad dips in the added-up  low-resolution BP/RP spectra even down to about 19$^{th}$ magnitude.

A very good overview on the impact of Gaia on the different subtopics dealing with white dwarfs is given in the white paper published by \cite {whitepaper} after the GREAT (Gaia Research for European Astronomy Training) 
workshop on "Gaia and the end states of stellar evolution" held at the University of Leicester in April 2011.
The science cases covered by the white paper are the mass-radius relation,
the initial-final mass relation,
the luminosity function,
magnetic fields in white dwarfs,
pulsating white dwarfs,
binaries with white dwarf components,
planetary systems around white dwarfs,
other late stages of stellar evolution,
fundamental physical parameters measured with white dwarfs,
hot subdwarfs,
and cataclysmic variables.

\section{Status of Gaia}
All 106 CCDs and the associated electronic modules are working well and are collecting data. The Gaia telescopes are focused, the orbit and attitude control is working very well, the satellite's spin rate is synchronised with the readout speed of the CCDs, the data collection hardware and software is fully operational, and the transmission of the data to the ground station works perfectly. 

The Astronomisches Rechen-Institut at the university of Heidelberg has the main responsibility for the so-called First Look of the Gaia mission. This is a very detailed daily analysis of the quality of the  
Gaia data including the One-Day Astrometric Solution ODAS. From the latter we could -- well before the first AGIS -- conclude that every single astrometric measurement of a 15$^{th}$ magnitude star on one CCD has a smaller noise in determining of the position than the end-of-mission accuracy of the HIPPACOS catalogue \citep {Hipparcoscatalogue,Hipparcosnew}; this is true even though we do not yet have a perfect calibration of the point-spead function (used to accurately determine the centroid of the stellar image) nor a chromaticity calibration.
These results of ODAS already show that a star catalogue of very high accuracy can - with enough data and a better understanding of the calibration -  later be produced by the global astrometric solution. 

Despite these very positive results  three serious issues have been discovered during the satellite's commissioning phase that may have impact on the performance of Gaia: 
\begin{itemize}
\item Firstly, it turned out that the routine week-long decontamination by active heating did not expel all residual water in the instrument part of the spaceraft. 
Water continues to steadily contaminate cold (less than -100$^\circ$ C) optical parts (e.g. the mirrors) resulting in a gradual loss of transmission of the optical system.  Therefore, three additional heating campaigns were necessary up to the end of September 2014 and probably several more will be necessary throughout the mission. While the decontamination is able to restore the full transmission of the optics, the necessary thermal and mechanical stability of Gaia is seriously interrupted by each such campaign.
\item Secondly, the basic angle between the two telescopes varies with an amplitude of about one milliarcsecond at the spin period of six hours, much larger than the few microarcseconds aimed for. Additionally, there is a significant day-to-day variation of the basic angle. The six-hour variation seems to be very stable but influences Gaia's ability to measure the zero point of the parallaxes without referring to far distant objects like quasars. 

If the basic-angle variation can be  measured accurately enough by the onboard interferometer (basic-angle monitor) and by the global astrometric solution  an appropriate calibration can minimize the negative effect on the accuracy of the positional measurements.

\item Thirdly, undesired stray light both from the astronomical sky and the Sun is entering the Gaia telescopes. 
The impact on the Gaia astrometry is small for objects brighter than magnitude 16 but increases towards lower flux levels. The RVS is more affected so that the faint limit needed to be  lowered by about one magnitude.
\end{itemize}

\section{Accuracy of the Gaia data}
A detailed description of the current expectation for the accuracy of Gaia data is given by \citet[][pre launch] {GaiaPerformance}  and on the Gaia website\footnote{http://www.cosmos.esa.int/web/gaia/science-performance}, taking into account all known instrumental effects, including the straylight levels as measured during the commissioning phase. 

For a 15$^{\rm th}$-magnitude star the predicted end-of-mission parallax standard errors remain close to the pre-launch value of 25 microarcseconds, slightly depending on the spectral type. The averaged position and proper-motion errors per year are of the same order. At 20$^{\rm th}$  magnitude the current estimations predict a standard error of 500-600 microarcseconds, about 50\%\ larger than estimated before the launch of Gaia.

There is also an assessment of the astrometric accuracy of bright stars ($3<V<12$) stating that a mean standard error of $5-14$ microarcseconds could be reached. However, it is currently difficult to predict how well image centroids for bright saturated stars can be determined.

All current estimations of the astrometric accuracy do not yet account for systematic errors caused by the basic-angle variation and thermal instabilities induced by the decontamination campaigns to re-establish full transmission of the Gaia optics. Such estimations will only be possible after the first AGIS, about 18 months after the start of the nominal mission.

The photometric standard errors in the integrated light, the BP, and the RP bands strongly depend on the spectral type and vary between $1-7$ millimagnitudes for a 15$^{\rm th}$ magnitude star and 
$6-490$ millimagnitudes for  a 20$^{\rm th}$ magnitude star. 

With the exception of white dwarfs with non-degenerate companions the RVS radial-velocity measurements are not relevant for white dwarfs research since no spectral lines are expected in the near-infrared region around the CaII doublet. The estimation of the radial-velocity error of non-degenerate stars varies from 1 to 15 kilometres per second depending on the brightness and spectral type.

\section{Release scenario}
There will be no proprietary data available for the Gaia team: All data releases will be available for everyone at the same time.  The final Gaia catalogue with astrometric, photometric, and spectroscopic results for all about one billion stars will not be ready before 2022. Besides the five-year measurement period of the Gaia a very detailed understanding of the measurements, an optimum reduction of the data, and a very sophisticated calibration and processing of the Gaia data is needed; these tasks will not be eased by complications induced by issues like the contamination, the basic-angle variation, and the additional straylight. 

Several intermediate releases are foreseen with limited samples and limited accuracy. These will be published as soon as  enough data are gathered, the necessary processing has been performed, and a reliable error estimation can be provided. 

On http://www.cosmos.esa.int/web/gaia/release the following data-release scenario is provided:
\begin{itemize}
\item {\bf Summer 2016:} Positions and magnitudes for all single stars with acceptable formal erors will be published. Additionally, a small special catalogue of objects within one degree distance from the ecliptical poles will be provided. For stars in common with the HIPPARCOS catalogue updated positions and proper motions will be released.
\item {\bf Early 2017:} Positions, proper motions and parallaxes of objects with single-star behaviour will be released together with integrated BP/RP photometry and mean radial velocities for a subset of objects.
\item {\bf 2017/2018:} First catalogue with  a subset of binary stars, object classifications, astrophysical parameters, and individual radial-velocity measurements.
\item {\bf 2017/2018:} First catalogue also containing a subset of variable-star classifications, non-single stars, and solar-system objects (asteroids, comets, planetary moons).
\item {\bf Final catalogue 2022:} Full astrometric, photometric, and spectroscopic catalogue with all single and multiple stars, source classifications, astrophysical parameters, as well as individual positional,  photometric and spectroscopic measurements. 
\end{itemize}

The Gaia team aims at releasing the intermediate catalogues at about the stated times, but this does not mean that these are fixed dates that one can rely on. 

\section{Conclusion}
Gaia is taking high-quality on a regular basis since end of July 2014. Throughout its five year nominal mission Gaia will probably discover several hundred thousand new white dwarfs providing astrometric, photometric, and (low-resolution) spectroscopic information 
down to 20$^{th}$ magnitude. In most cases the newly discovered white dwarf will need follow-up observations for a detailed astrophysical analysis. 

For the known white dwarfs in the magnitude range between 12 and 20 a standard error in position, proper motion per year, and parallaxes  between several dozen and a few hundred microarcsecond can be expected  for the final catalogue
The situation for brighter white dwarfs or brighter companions of white dwarfs is less certain. Moreover, 
one has to wait for the first global astrometric solutions in oder to assess the size of systematic effects caused e.g. by the basic-angle variation.

In order to make best use of the Gaia results, everyone should be prepared to have the necessary tools to work with the Gaia data available before the intermediate and final catalogues will be published. It will be possible to perform pre-release tests with mock-up catalogues. A simulated Gaia catalogue and the initial Gaia source list is available on an ESA website \footnote{https://geadev.esac.esa.int/gacs-dev/} via an SQL query form.


\bibliography{Gaia_WD}  

\end{document}